\documentclass{PoS}

\usepackage{bm,amsmath,amssymb}
\usepackage[mathscr]{eucal}

\long\def\comment#1{ }

\newcommand{\beq}{\begin{equation}}
\newcommand{\eeq}{\end{equation}}
\newcommand{\bea}{\begin{eqnarray}}
\newcommand{\eea}{\end{eqnarray}}

\newcommand{\dif}{{\rm d}}

\newcommand{\rmi}{{\rm i}}
\newcommand{\rmP}{{\rm P}}

\newcommand{\wh}{\widehat}

\newcommand{\bx}{\bm{x}}

\newcommand{\abar}{\bar{\alpha}_s}

\newcommand{\deltam}{\delta\hspace{-.025cm}}

\def\1eq#1{Eq.~(\ref{#1})}

\def\2eqs#1#2{Eqs.~(\ref{#1}) and~(\ref{#2})}
\def\3eqs#1#2#3{Eqs.~(\ref{#1}),~(\ref{#2}) and~(\ref{#3})}
\def\noeq#1{(\ref{#1})}

\def\cso{\delta c}
\def\bcso{\delta \bar c}
\def\bso{\delta b}

\def\G{\Gamma}

\title{High-energy QCD evolution\\ from Slavnov-Taylor identity}

\ShortTitle{High-energy QCD evolution from ST identity}

\author{\speaker{Andrea Quadri}\\
        INFN, Sez. di Milano and Dip. di Fisica, Univ. di Milano\\
        via Celoria 16, I-20133 Milan\\
        Italy\\
        E-mail: \email{andrea.quadri@mi.infn.it}}

\abstract{We clarify the derivation of high-energy QCD evolution equations
 from the fundamental gauge symmetry of QCD.
The gauge-fixed classical action of the Color Glass Condensate (CGC)
is shown to be invariant
under a suitable BRST symmetry, that holds after the separation of the
gluon modes into their fast classical (background) part, the soft
component and the semifast one, over which the one-step
quantum evolution is carried out.
The resulting Slavnov-Taylor (ST) identity holds to all orders in perturbation theory
and strongly constrains the CGC effective field theory (EFT) arising from
the integration of the soft modes. 
We show that the ST identity guarantees gauge-invariance of the
EFT. It also allows to control the dependence on the gauge-fixing
choice for the semifast modes (usually the lightcone gauge
in explicit computations). The formal properties of the evolution
equations valid in different regimes (BKFL, JIMWLK, ...) can be
all derived in a unified setting within this algebraic approach.}

\FullConference{QCD-TNT-III-From quarks and gluons to hadronic matter: A bridge too far?,\\
		2-6 September, 2013\\
		European Centre for Theoretical Studies in Nuclear Physics and Related Areas (ECT*), Villazzano, Trento (Italy)}

\begin{document}

\section{Introduction}

The Color Glass Condensate (CGC)~\cite{Gelis:2010nm} is an effective field theory
approach to the physics of high gluon densities and gluon saturation.
The fast gluon modes (in the infinite momentum frame) are described
by the classical solution to the Yang-Mills equation in the presence
of some static color sources $\rho$. The latter are associated with a
weight function $W_\Lambda[\rho]$, characterizing the CGC at the longitudinal
scale $\Lambda$. 
Quantum QCD dynamics introduces radiative corrections around the
classical (background) gluon configuration. These corrections
affect the updated weight function $W_{b\Lambda}[\rho]$ at the new
scale $b \Lambda$, with $b \ll 1$. Even when 
$\abar \ln 1/b \ll 1$ (with $\abar = \alpha_x N/\pi$,
$N$ being the number of colors), so that perturbation theory can be trusted,
the quantum corrections give rise to logarithmically enhanced contribution that must be resummed
thorugh renormalization-group (RG) techniques. 
Depending on the different approximations used, one ends up with
the well-known BFKL~\cite{Lipatov:1976zz}-\cite{Balitsky:1978ic} and JIMWLK evolution equations~\cite{JalilianMarian:1997jx}-\cite{Weigert:2000gi}.

In this proceeding I would like to provide an introduction to the recent work
aimed at clarifying the role of fundamental QCD gauge symmetry in
constraining the properties of the effective field theory (EFT) of the CGC.
The motivation is to separate the general features of the evolution equations, 
that only depend on the symmetry content of the model, from the
specific aspects related e.g. to the choice of the gauge for
the semifast modes or the particular approximation used
in the computation of the EFT.

It turns out that such a program is indeed quite successful.
The main results are the following. 
Gauge-invariance of the EFT (after the one-step
quantum evolution) can be proven on the basis of the Slavnov-Taylor 
 (ST) identity only. The ST identity encodes at the quantum level
the classical BRST invariance of the QCD action.
The ST identity also guarantees that the classical Yang-Mills
equations of motion are not deformed by the quantum corrections. Thus
the classical description of the CGC at the new scale $b\Lambda$ 
in terms of a modified 
weight function $W_{b\Lambda}[\rho]$, with the {\em same} equations
of motion holding at the scale $\Lambda$,
is indeed consistent. 
This is a crucial ingredient in the derivation of the evolution
equations and is far from being obvious. It ultimately
relies on the peculiar BRST symmetry of the CGC theory, after the
separation of the gluons into their components (fast, semifast and soft ones).

The technical tools required for this program have been recently developed
in a series of papers~\cite{Binosi:2011ar}-\cite{Binosi:2012st} where cohomological methods and generalized 
Lie series techniques have been used in order
to constrain the complete background dependence in a gauge theory (for the full
vertex functional and not only for its local approximation). 

We will not try to sketch the proofs of the results 
for the CGC EFT here. Details can be
found in a forthcoming paper~\cite{Binosi:2014xua}.

\section{Classical Action}

The classical action of the CGC is
\beq
S_{CGC}[A,\rho] = S_{YM}[A]+ S_W[A,\rho]
\eeq
where we denote by $S_{YM}$ the $SU(N)$ Yang-Mills action
\beq
S_{YM} = - \int d^4x \, \frac{1}{N} Tr[F_{\mu\nu} F^{\mu\nu}]
\eeq
with $F_{\mu\nu} = F^a_{\mu \nu} T^a$ and $T^a$ the $SU(N)$ generators.
In components the field strength reads
$F^a_{\mu\nu} = \partial_\mu A^a_\nu - \partial_\nu A^a_\mu +
g f^{abc} A^b_\mu A^c_\nu$. $f^{abc}$ are the $SU(N)$ structure constants.

It is convenient to make use of the light-cone coordinates
$x^\mu = (x^+,x^-,{\bf x})$ with $x^\pm = (x^0 \pm x^3)/\sqrt{2}$
and ${\bf x} = (x^1,x^2)$. We also use the vector notation
$\vec{x} = ( x^-, {\bf x})$.

$\rho$ is the plus component (the only non-vanishing one) of the
color current associated with the fast sources. In the infinite
momentum frame $\rho=\rho(\vec{x})$ is static.
The interaction with the gluon field is contained in $S_W$.
Under the assumption that in lowest order the coupling is proportional to
$\rho_a(\vec{x}) A^-_a(x)$ and by requiring gauge-invariance,
$S_W$ turns out to be defined on some suitable  Schwinger-Keldysh contour $C$
in the complex time plane,
built by joining the path on the real axis from $-\infty$ to some $x_f^+$ and backwards
on a path from $x_f^+$ to $-\infty$ with a small imaginary part and afterwards by taking the limit
$x_f \rightarrow + \infty$.
$S_W$ is given by~\cite{Ferreiro:2001qy}
\beq
S_W[A,\rho] = \frac{i}{gN} \int d^3 \vec{x} ~ {\rm Tr} [p(\vec{x}) W_C(\vec{x})] \, ,
\eeq
where
$W_C(\vec{x})$ is the contour temporal Wilson line
\beq
W_C(\vec{x}) = T_C \exp \Big [ i g \int_C dz ~ A^-(z,\vec{x}) \Big ] \, .
\eeq
In the infinite momentum frame (IMF) one can separate
the fast gluon modes form the soft ones, i.e. we  set
 \beq
 A_{\mu} = \wh{A}_{\mu} + a_{\mu} + \deltam A_{\mu} .
 \label{gluonsplit} 
\eeq
$\wh{A}_{\mu}$ represents the fast modes 
with longitudinal momenta $|p^+| > \Lambda$.
It is fixed by the classical solution 
of the equation of motion of the CGC action $S_{CGC}$.
$a_{\mu}$ describes the semi-fast modes with momenta $p^+$ 
such that $\Lambda > |p^+| > b \Lambda$ (where we assume that 
$b \ll 1$ but with $\abar \ln(1/b) \ll 1$). The $a$-modes are the quantum
fields that will be integrated out during the one-step quantum evolution.
$\delta A_{\mu}$ are in turn the soft modes with momenta 
$|p^+| < b \Lambda$.
They are fixed configurations during the one-step quantum evolution,
so that the background actually has two components, a fast one
($\wh{A}_\mu$) and a soft one ($\delta A_{\mu}$).

\section{BRST Symmetry}

The BRST symmetry of the full gluon field $A^a_\mu$ is obtained
by replacing the gauge parameter with the Faddeev-Popov ghost field $C^a$,
namely
\beq
s A^a_\mu = D^{ab}_\mu[A]C^b \, , \qquad D^{ab}_\mu[A]=\delta^{ab} \partial_\mu + 
g f^{acb} A^c_\mu 
\label{brst.gluon}
\eeq
The background $\wh{A}^a_\mu$ is paired into a BRST doublet~\cite{Quadri:2002nh},
\cite{Barnich:2000zw} through
\beq
s \wh{A}^a_\mu = \Omega^a_\mu \, , \qquad s \Omega^a_\mu = 0 \, .
\label{brst.bkg}
\eeq
$\Omega^a_\mu$ is an anticommuting external source with ghost
number one.
Eq.(\ref{brst.bkg}) guarantees that the cohomology of the theory~\cite{Barnich:2000zw}
(and hence the physical observables) is unaffected
by the introduction of the background configuration.
In the standard approach to the algebraic
treatment of the Background Field Method
\cite{Grassi:1995wr}-\cite{Becchi:1999ir},
Eq.(\ref{brst.gluon}) together with eq.(\ref{brst.bkg})
fixes uniquely the BRST transformation of the quantum field $Q_\mu=A_\mu-\wh{A}_\mu$.
In the present case there is one more field in the game, namely $\delta A$.
However in the CGC prescription for obtaining the EFT $\delta A$ plays the role of a background field and thus cannot have a $\Omega$-dependent term
in its BRST variation.
It is also convenient to split the ghost $C^a$ into a soft ($\delta c^a$) and a semifast ($c^a$) component according to
\beq
C^a = c^a + \delta c^a
\eeq
This leads us to the following BRST transformations for the gluon modes:
\beq
s \delta A^a_\mu = D^{ab}_\mu[\delta A] \delta c^b \, , \qquad
s a^a_\mu = s A^a_\mu = s \delta A^a_\mu - s \wh{A}^a_\mu \, .
\label{brst.gauge}
\eeq

\medskip
In the presence of a soft component of the gluon field that is not
integrated out the color current entering into $S_W$
 must be evaluated by using the
Wilson line from $z^+ \rightarrow -\infty$ to $x^+$ of the soft modes
$\delta A^-$~\cite{Binosi:2014xua}.
The classical equation of motion then becomes 
(we denote by $\alpha$ the plus component of the fast background field,
that can be chosen to be the only non-vanishing one in the Coulomb gauge):
 \beq
 \bm{\nabla}_{\bx}^2 \alpha(x) = - U^\dagger(x) J^+(x) U(x)
  \label{eom}
 \eeq
where $J^+$ is the plus component of the color-rotated current and 
$U^\dagger$ is the Wilson line
 \beq
 \label{wilsonu}
 U^{\dagger}(x) = 
 \rmP \exp \left[ \rmi g \int_{-\infty}^{x^-}
 \dif z^- \alpha(x^+,z^-,\bx) 
 \right], 
 \eeq
It is convenient to redefine the source in the r.h.s. of
eq.(\ref{eom}) by setting
\beq
\label{chi-charge}
\chi(x) \equiv  U^\dagger(x) J^+(x) U(x).
\eeq
Then, $\chi$ becomes the independent variable.
 $\chi$ transforms in the adjoint representation of SU(N), and therefore
\beq
\label{BRST.chi}
s \chi^a = g f^{abc} \chi^b \cso^c.
\eeq
Again one has to split $\chi$ into a background part and a quantum
fluctuation
\beq
\chi^a = \widehat{\chi}^a + \delta \chi^a \, .
\label{chi.split}
\eeq
$\widehat{\chi}^a$ is the source associated to the classical field
$\alpha$ and hence its BRST transformation is given by
\beq
\label{sonwhchi}
s \wh{\chi}^a = - \bm{\nabla}_{\bx}^2 \Omega^{a+}(x).
\eeq
while the BRST transformation of $\delta \chi$ is obtained by
difference from Eq.(\ref{BRST.chi}) and Eq.(\ref{chi.split}).

\section{ST identity for the EFT}

The quantization of the theory proceeds by choosing a gauge-fixing
for the semifast modes. This is usually done in the lightcone
gauge, where ghosts decouple. This is why they never appear
in explicit computations in the CGC framework. 
 However, having introduced the appropriate BRST symmetry, nothing
prevents to choose a different gauge. The ghost interactions
are automatically fixed by the BRST symmetry in the usual
fashion and one must of course include
their contributions in the computation of
the effective action at leading order 
in $\abar \ln(1/b)$. This is discussed in detail in~\cite{Binosi:2014xua}.

The existence of a classical BRST-invariant action allows one to write
in the  usual way~\cite{Gomis:1994he} the Slavnov-Taylor (ST) identity
for the full vertex functional $\G$, valid to all orders in the loop
expansion if the theory is non-anomalous (as it happens for QCD).

In the CGC framework one is not really interested in $\G$, but in
the effective action $\widetilde \G$ obtained by intgrating out the semifast modes 
$a_\mu$. $\widetilde \G$ is the generating functional of all diagrams
that are one-particle reducible (1-PR) with respect to (w.r.t.)
$a_\mu$ and one-particle irreducibile (1-PI) w.r.t. all other fields.
We will not dwell here on the details of the derivation of $\tilde \G$
and just report the ST identity that it fulfills:
\begin{align}
{\cal S}\hspace{0.025cm}\widetilde \Gamma\equiv\int\!\dif^4z&\left[\Omega^a_\mu(z)\frac{\delta\widetilde \Gamma}{\deltam\widehat{A}^a_\mu(z)}+\frac{\delta\widetilde \Gamma}{\delta(\deltam{A}^{*a}_\mu(z))}\frac{\delta\widetilde \Gamma}{\delta(\deltam{A}^a_\mu(z))}
+\frac{\delta\widetilde \Gamma}{\delta(\cso^{*a}(z))}\frac{\delta\widetilde \Gamma}{\delta(\cso^a(z))}+\bso^a(z)\frac{\delta\widetilde \Gamma}{\delta(\bcso^a(z))}\right.\nonumber \\
&\left.+\frac{\delta\widetilde \Gamma}{\delta(\deltam{\chi}^{*a}(z))}\frac{\delta\widetilde \Gamma}{\delta(\deltam{\chi}^a(z))}\right]=0.
\label{ST-tildeGamma}
\end{align}
$\delta \bar c^a$ is the antighost for the soft modes and
$\delta b^a$ the corresponding Nakanishi-Lautrup multiplier field.
$\delta A^*$, $\delta c^*$ and $\delta \chi^*$ are the antifields (i.e. the sources
of the BRST transformations) of the soft gluons, the soft ghost and
the quantum fluctuations of the color sources $\chi$ respectively.

\subsection{Gauge invariance}

Eq.~\noeq{ST-tildeGamma} yields very strong constraints on
the effective action of the CGC. Let us 
take a derivative w.r.t. $\cso^b$ and
then set $\cso$, $\Omega$ and $\bso$, to zero.
One finds
\beq
 \int \dif^4z \left[ 
\frac{\delta^2 \widetilde \Gamma}{\delta (\cso_b(x)) 
\delta(\deltam{A}^{*a}_\mu(z))} \frac{\delta\widetilde \Gamma}{\delta(\deltam{A}^a_\mu(z))}
+
\frac{\delta^2 \widetilde \Gamma}{\delta (\cso_b(x))
\delta(\deltam{\chi}^{*a}_\mu(z))} \frac{\delta\widetilde \Gamma}{\delta(\deltam{\chi}^a_\mu(z))} \right]=0.
\label{ward.id}
\eeq
At $\Omega=0$ the BRST variations  $s\, \delta A$
and $s\, \delta \chi$ contain neither $a$ nor $c$ and hence
do not receive radiative corrections upon the integration of the
semifast modes.
This means that they remain classical, namely
\begin{align}
& \frac{\delta^2 \widetilde \Gamma}{\delta (\cso_b(x)) 
\delta(\deltam{A}^{*a}_\mu(z))} = \frac{\delta^2 \widetilde \Gamma^{(0)}}{\delta (\cso_b(x)) 
\delta(\deltam{A}^{*a}_\mu(z))} =  \delta^{ab} \partial_\mu \delta^{(4)}(x-z)
+ g f^{acb} \delta A^c_\mu \delta^{(4)}(x-z), \nonumber \\
& \left . \frac{\delta^2 \widetilde \Gamma}{\delta (\cso_b(x))
\delta(\deltam{\chi}^{*a}(z))} \right |_{\Omega = 0} =
 \left . \frac{\delta^2 \widetilde \Gamma^{(0)}}{\delta (\cso_b(x))
\delta(\deltam{\chi}^{*a}(z))} \right |_{\Omega = 0} =
g f^{acb} \left[\wh\chi^c(x) + \delta \chi^c(x)\right] \delta^{(4)}(x-z).
\end{align}
Hence \1eq{ward.id} amounts to the statement
of gauge invariance of the effective action $\widetilde \Gamma$.
We stress that this result holds
{\em irrespectively of the gauge choice for the semi-fast modes}: that is, gauge invariance follows as a consequence of the ST identity
(\ref{ST-tildeGamma}) and of the 
semifast-soft decomposition.

\section{(Quantum-deformed) Background Equations of Motion}

Taking a functional differentiation of~\1eq{ST-tildeGamma} w.r.t. the source~$\Omega$, and setting $\Omega=0$ afterwards, yields (at $\delta b=0$)
\beq
\frac{\delta\widetilde\Gamma}{\deltam\widehat{A}^a_\mu(x)}=-\int\!\dif^4z\left[\frac{\delta\widetilde\Gamma}{\delta\Omega^{a}_\mu(x)\delta(\deltam{A}^{*b}_\nu(z))}\frac{\delta\widetilde\Gamma}{\delta(\deltam{A}^a_\mu(z))}
+\frac{\delta\widetilde\Gamma}{\delta\Omega^{a}_\mu(x)\delta(\deltam{\chi}^{*b}(z))}\frac{\delta\widetilde\Gamma}{\delta(\deltam{\chi}^b(z))}
\right].
\label{EoM-fundam}
\eeq
This is a fundamental equation for the effective field theory, as it encodes how quantum fluctuations will modify the classical equation of motion~\noeq{eom} for the background field configuration.

The first term in~\noeq{EoM-fundam} controls the (gauge-dependent) deformation of the classical background-quantum splitting induced by quantum corrections~\cite{Binosi:2011ar,Binosi:2012pd,Binosi:2012st}.
In general (e.g. for gauge theories in the presence of instanton configurations) this deformation is not zero.

The second term fixes instead the functional dependence of the background $\wh{A}$ on the color charge density $\deltam\chi$, once quantum corrections are taken into account. 

This result is very general: it does not rely on the specific
form of the action chosen, on the gauge-fixing adopted for the
semifast modes, and on the order of approximation used in order to compute
radiative corrections.

Let us now specialize to the CGC case. By taking into account the 
structure of the BRST symmetry it can be proven~\cite{Binosi:2014xua}
that
\beq
\widetilde{\Gamma}_{\Omega^\mu_a\deltam A^{*\nu}_b}(x,y)=0,
\label{r1}
\eeq
while
\begin{align}
&\widetilde{\Gamma}_{\Omega^\mu_a \delta\chi^*_b}(x,y)  = \Gamma^{(0)}_{\Omega^\mu_a \delta \chi^*_b}(x,y) = - \delta^{\mu+}\bm \nabla^2_{\bx} \delta^{(4)}(x-y).
\label{r3}
\end{align}
It is understood that the background field has not been set to zero here,
so Eqs.(\ref{r1}) and (\ref{r3}) are valid in the presence of the background $\alpha$.
Eq.(\ref{r1}) tells us that the equation of motion for the background
will not be affected by the soft modes of the theory. This is physically
reassuring, since it means that the separation of scales is preserved under
the quantum evolution. 
Eq.(\ref{r3}) in turn implies that the equation of motion, relating the
background field to the color sources, maintains the same {\em classical} form
after quantum corrections are taken into account.
This is a non-trivial result that follows directly from the symmetry
properties of the theory. 
It is a crucial property in order to be able to encode all the effects of the quantum fluctuations into an updated weight function $W_{b\Lambda}[\rho]$, while
determining the new background configuration again by the same
classical equation of motion valid for the theory at the scale $\Lambda$.

\section{Conclusions}

Gauge invariance of the CGC effective action, as well as 
the stability of  the background equation of motion
under the one-step quantum evolution,
follow from basic symmetry properties of the theory, encoded
in the ST identity. 
In particular it does not depend on the choice of the
semifast modes and on the order of approximation used
in the evaluation of the effective action.
The general structure of the evolution equations can also be
derived within this purely algebraic setting~\cite{Binosi:2014xua}.

\appendix
\section*{Acknowledgments}

The author wishes to acknwoledge the warm hospitality and the
support of ECT* during the workshop.

\end{document}